\documentclass[prd,preprint,superscriptaddress,amsmath,amssymb,nofootinbib]{revtex4}
\usepackage{graphicx}
\usepackage{dcolumn}
\usepackage{bm}
\usepackage{amssymb}
\usepackage{amsmath}
\usepackage{epsfig}    
\usepackage{color}
\usepackage{slashed}
\usepackage{hhline}

\def\be{\begin{equation}}
\def\ee{\end{equation}}
\newcommand{\bea}{\begin{eqnarray}}
\newcommand{\eea}{\end{eqnarray}}
\newcommand{\nn}{\nonumber}

\begin{document}

 \begin{flushright} {KIAS-P18093}, APCTP Pre2018 - 014  \end{flushright}

\title{Inverse seesaw model with large $SU(2)_L$ multiplets \\
and natural mass hierarchy}

\author{Takaaki Nomura}
\email{nomura@kias.re.kr}
\affiliation{School of Physics, KIAS, Seoul 02455, Republic of Korea}

\author{Hiroshi Okada}
\email{okada.hiroshi@apctp.org}
\affiliation{Asia Pacific Center for Theoretical Physics, Pohang, Geyoengbuk 790-784, Republic of Korea}

\date{\today}

\begin{abstract}
We propose an inverse seesaw model with large $SU(2)_L$ multiplet fields which realizes natural mass hierarchies among neutral fermions.
Here, lighter neutral fermion mass matrices are induced via two suppression mechanisms; one is small vacuum expectation value of $SU(2)_L$ triplet required by rho parameter constraint and the other is generation of Majorana mass term of extra singlet fermions at one-loop level. To realize the loop masses, we impose $Z_2$ symmetry which also guarantees stability of a dark matter candidate.
Furthermore, we discuss anomalous magnetic moment and collider physics from interactions of large multiplet fields.
 \end{abstract}
\maketitle

\section{Introductions}
A model of inverse seesaw~\cite{Mohapatra:1986bd, Wyler:1982dd} or linear seesaw~\cite{Wyler:1982dd, Akhmedov:1995ip, Akhmedov:1995vm} is well-known as one of the elegant mechanisms to generate Majorana masses for active neutrinos, including both the left and right handed heavier neutral fermions.
Thus, it is frequently discussed in a larger gauge theory such as $SU(2)_L\times SU(2)_R$~\cite{Mohapatra:1979ia}, $( SU(2)_L\times SU(2)_R\subset) SO(10)$~\cite{so10}, {SO(10) with supersymmetry~\cite{Malinsky:2005bi}} etc, as an unified theory. 
On the other hand, there are several issues which should be improved in these models.
A representative issue is how to describe more natural hierarchies among neutral fermion mass matrices.
For example we need to assume small Majorana mass for singlet fermion in realizing inverse seesaw mechanism unless there is a way to justify the smallness. 
In addition, small Yukawa couplings are required to obtain Dirac mass term among SM neutrinos and singlet fermion to fit neutrino oscillation data with TeV scale heavy neutrinos.
It could not be well explained by simple gauge extended scenarios of the standard model (SM)~\footnote{In ref.~\cite{Bazzocchi:2009kc}, small Majorana mass in inverse seesaw mechanism is induced by a small VEV generated by supersymmetry breaking renormalization group equation effect in a supersymmetric model. }.

In this letter, we realize the inverse seesaw mechanism of natural hierarchies among neutral fermion mass terms by introducing extra fields with larger $SU(2)_L$ representations and applying radiatively induced mass mechanism for Majorana mass term. In order purely to realize inverse seesaw,
we impose a global $U(1)$ lepton number that forbids linear seesaw.
A scalar field with larger $SU(2)_L$ representations restricts their vacuum expectation values (VEVs) to be or less than the order of 1 GeV by constraint from $\rho$-parameter~\cite{Nomura:2018lsx, Nomura:2018cfu, Nomura:2018ibs, Nomura:2018cle, Nomura:2017abu,Nomura:2016jnl, Nomura:2016dnf}.
In this model $SU(2)_L$ triplet scalar develops a VEV which provides Dirac mass among the SM neutrino and singlet fermion which are typically 1 GeV or less 
because of the restricted VEV.
The Majorana mass term of singlet fermion is generated at one loop level providing an additional loop suppression factor~\cite{Ma:2006km}
where we impose a $Z_2$ symmetry to realize the mechanism.
As a bonus of this additional symmetry, we  have a dark matter (DM) candidate.  
Furthermore, we also explain anomalous magnetic moment of muon and discuss collider physics focusing on interactions of large multiplet fields.

This letter is organized as follows.
In Sec. II, {we review our model and formulate the Yukawa sector and Higgs sector, remormalization group equations (RGEs),
lepton flavor violations (LFVs), muon anomalous magnetic moment, and DM candidate.
In Sec. III, we show numerical analysis at a benchmark point including all the constraints and discuss collider physics.
Finally we devote the summary of our results and the conclusion.}

\section{Model setup and Constraints}
\begin{table}[t!]
\begin{tabular}{|c||c|c|c|c||c|c|c|c|}\hline\hline  
& ~$L_L^a$~& ~$e_R^a$~& ~$\psi^a$~& ~$\chi_R^a$~& ~$H_2$~& ~$H_3$~& ~$H_4$~& ~$\varphi$~ \\\hline\hline 
$SU(2)_L$   & $\bm{2}$  & $\bm{1}$  & $\bm{4}$  & $\bm{1}$ & $\bm{2}$ & $\bm{3}$  & $\bm{4}$  & $\bm{1}$   \\\hline 
$U(1)_Y$    & $-\frac12$  & $-1$ & $-\frac12$  & $0$  & ${\frac12}$ & $0$  & $\frac12$    & $0$ \\\hline
$U(1)_L$    & $\ell$  & $\ell$ & $\ell$  & $\ell$  & $0$  & $0$  & $0$    & $-2 \ell$ \\\hline
$Z_2$    & $+$  & $+$ & $+$  & $-$  & $+$  & $+$  & $-$  & $+$ \\\hline
\end{tabular}
\caption{Charge assignments of the our lepton and scalar fields
under $SU(2)_L\times U(1)_Y\times U(1)_L\times Z_2$ with $\ell\neq 0$, where the upper index $a$ is the number of family that runs over 1-3,
all of them are singlet under $SU(3)_C$, and the quark sector is same as the SM one.}\label{tab:1}
\end{table}

In this section we introduce our model.
First of all, we introduce a global $U(1)_L$ symmetry to obtain a successful inverse seesaw model, which will be spontaneously broken. 
Also we impose $Z_2$ symmetry in order to realize natural hierarchies among neutral fermions and assure a stable DM candidate, as we will discuss below.
As for the fermion sector, we introduce three families of vector fermions $\psi$ with $(4,{-1/2,\ell,+})$, and right-handed fermions $\chi_R$ with $(1,0,\ell,-)$, where each of content in parentheses represents the charge assignment of ($SU(2)_L, U(1)_Y, U(1)_L, Z_2$) symmetry.
As for the scalar sector, we add a triplet scalar field $H_3$ with $(3,{0,0,+})$, a quartet inert scalar field $H_4$  with $(4,{1/2,0,-})$, and a singlet scalar field $\varphi$  with $(1,{0,-2\ell,+})$, where SM-like Higgs field is denoted as $H_2$.
Here we write vacuum expectation values {(VEVs)} of $H_{2,3}$ and $\varphi$ by $\langle H_{2,3}\rangle\equiv v_{2,3}/\sqrt2$ and $\langle \varphi\rangle\equiv v_\varphi/\sqrt2$
which induces the spontaneously electroweak and $U(1)_L$ symmetry breaking.
All the field contents and their assignments are summarized in Table~\ref{tab:1}, where the quark sector is exactly the same as the SM.
The renormalizable Yukawa Lagrangian under these symmetries is given by
\begin{align}
-{\cal L_\ell}
& =  y_{\ell_{aa}} \bar L^a_L H_2 e^a_R  +  y_{D_{ab}} [ \bar L^a_L H_3 \psi_R^b ]  
 +  g_{ab} [\bar \psi^{a}_L  H_4^* \chi^b_R]
 \nn\\
&+M_{{aa}} \bar \psi^a_L \psi_R^a+
\frac{y_{\varphi_{aa}}}2 \varphi (\bar \chi^c_R)^a \chi^{a}_R
+ {\rm h.c.}, \label{Eq:yuk}
\end{align}
where we implicitly symbolize the gauge invariant contracts of $SU(2)_L$ index as bracket [$\cdots$] hereafter,
indices $(a,b)=1$-$3$ are the number of families, and ($y_\ell,M_D, y_{\varphi}$) are assumed to be  diagonal matrix with real parameters
without loss of generality. Then, the mass eigenvalues of charged-lepton are defined by $m_\ell=y_\ell v/\sqrt2={\rm Diag}(m_e,m_\nu,m_\tau)$. 
Note here that we do not have $[(\bar\psi^c_R)^a  H_4\chi^b_R]$ operator thanks to the global $U(1)_L$ symmetry, and thus the neutrino mass matrix can dominantly be induced via inverse seesaw.

\noindent \underline{\it Scalar potential and VEVs}:
The scalar potential in our model is given by
\begin{align}
{\cal V} = & -\mu_H^2 |H_2|^2  + \frac{\lambda_H}{2} |H_2|^4 + M^2_3 H_3^\dagger H_3 + M_4^2 H_4^\dagger H_4 - \mu_\varphi^2 \varphi^* \varphi + \frac{\lambda_\varphi}{2} (\varphi^* \varphi)^2 + \lambda_{H\varphi} (H^\dagger H)(\varphi^* \varphi) \nn\\
&+\mu_{32} (H^\dag_2H_3H_2+{\rm h.c.}) + \sum_i \lambda_{H_4 H_2}^i [H_4^\dag H_2 H_4^\dag H_2 + h.c.]_i 
+ ({\rm other \ trivial\ terms}),
\label{Eq:potential}
\end{align}
where sum for $(H^\dagger_4 H_2)^2$ term is for independent contraction patterns of $SU(2)_L$ index.
Non-zero VEVs of scalar fields are obtained by solving the conditions
\begin{equation}
\frac{\partial V}{\partial v_2} = \frac{\partial V}{\partial v_3} = \frac{\partial V}{\partial v_\varphi} = 0,
\end{equation}
where we assume VEV of $H_4$ to be  zero.
Taking condition $v_3 \ll v_2$ as we see below, the VEVs are approximately given by
\begin{equation}
v_2 \simeq \sqrt{\frac{\mu_H^2}{\lambda_H}}, \quad v_3 \simeq \frac{\mu_{32} v^2_2}{M_3^2}, \quad v_\varphi \simeq \sqrt{\frac{\mu_\varphi^2}{\lambda_{\varphi}}},
\end{equation}
where we ignored contributions from trivial terms in the potential like $(H^\dagger H) \varphi^* \varphi$ assuming their couplings are small.
The SM Higgs VEV is identified as $v_2 \sim 246$ GeV. 

\noindent \underline{\it $\rho$ parameter}:
The electroweak $\rho$ parameter deviates from unity due to the nonzero value of $v_3$ at the tree level as
\begin{align}
\rho=\frac{v_2^2+4v_3^2}{v_2^2},
\end{align}
where the experimental bound is $\rho_{\rm exp}=1.0004^{+0.0003}_{-0.0004}$ at 2$\sigma$ C.L.~\cite{Agashe:2014kda}. 
It suggests that 
\begin{align}
v_3\lesssim 3.25\ {\rm GeV},
\end{align}
where $\sqrt{v_2^2+v_3^2}\approx$246 GeV.

\noindent \underline{\it Exotic particles} :
The scalars and fermions with large $SU(2)_L$ multiplet provide exotic charged particles.
Here we write components of multiplets as
\begin{align}
& H_3 = (\delta^{+}, \delta^{0}, \delta^{'-})_R^T,\label{eq:H3}  \\
& H_4 = (\phi_4^{++}, \phi_4^{+}, \phi_4^{0}, \phi^{'-}_4)^T,  \label{eq:H4} \\
& \psi_{L(R)} = (\psi^{+}, \psi^{0}, \psi'^{-}, \psi^{--})^T_{L(R)} \label{eq:psiLR}. 
\end{align}
 The mass of component in $H_4$ and $\psi$ are given by $\sim M_4$ and $\sim M$ respectively, where charged particles in the same multiplet have degenerate mass at tree level which will be shifted at loop level~\cite{Cirelli:2005uq}.  Components of $H_3$ have also degenerated mass of $\sim M_3$ since VEV $v_3$ is small and masses are not much shifted.
We also have extra singlet scalar $\varphi$ which is written by
\begin{equation}
\varphi = \frac{1}{\sqrt{2}} (v_\varphi + \tilde \phi + i a_\varphi),
\end{equation}
where $a_\varphi$ is identified as light Goldstone boson associated with global lepton number symmetry breaking.
CP-even component $\tilde \phi$ can mix with the SM Higgs and mass term becomes 
\begin{equation}
\mathcal{L} \supset \frac{1}{4} \begin{pmatrix} \tilde H \\ \tilde \phi \end{pmatrix}^T \begin{pmatrix} \lambda_H v^2 & \lambda_{H \varphi} v v_\varphi \\  \lambda_{H \varphi} v v_\varphi  & \lambda_\varphi v_\varphi^2 \end{pmatrix} \begin{pmatrix} \tilde H \\ \tilde \phi \end{pmatrix},
\end{equation} 
where $\tilde H$ is neutral CP-even component in Higgs doublet $H_2$.
This squared mass matrix can be diagonalized by an orthogonal matrix and the mass eigenvalues are given by
\begin{equation}
m_{h,\phi}^2 = \frac{\lambda_H v^2 +\lambda_\varphi v_\varphi^2 }{4} \pm \frac{1}{4} \sqrt{\left( \lambda_H v^2 -\lambda_\varphi v_\varphi^2 \right)^2 + 4 \lambda_{H \varphi}^2 v^2 v_\varphi^2 }. 
\end{equation}
The corresponding mass eigenstates $h$ and $\phi$ are obtained as   
\begin{equation}
\begin{pmatrix} h \\ \phi \end{pmatrix} = \begin{pmatrix} \cos \alpha & \sin \alpha \\ - \sin \alpha & \cos \alpha \end{pmatrix} \begin{pmatrix} \tilde H \\ \tilde \phi \end{pmatrix}, \quad
\tan 2 \alpha = \frac{2 \lambda_{H \varphi} v v_\varphi}{\lambda_H v^2 - \lambda_\varphi v_\varphi^2},
\label{eq:scalar-mass-fields}
\end{equation}
where $\alpha$ is the mixing angle and $h$ is the SM-like Higgs boson with $m_h \simeq 125$ GeV. 

\begin{figure}[tb]
\begin{center}
\includegraphics[width=10.0cm]{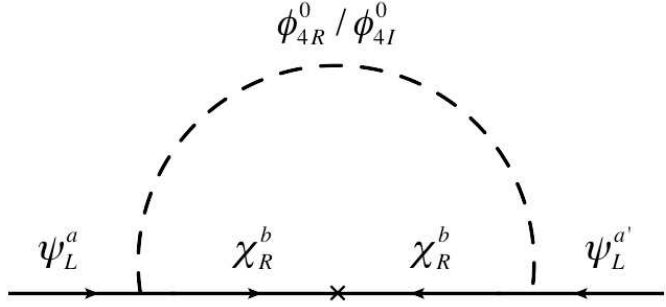}
\caption{Feynman diagram to generate the masses of $\mu_{L}$.}
\label{fig:mu_mass}
\end{center}\end{figure}

\subsection{Neutral fermion masses}
\noindent \underline{\it  Neutral sector}:
After the spontaneous symmetry breaking, neutral fermion mass matrix in basis of $\Psi^0_L\equiv (\nu_L, \psi^c_R,\psi_L^{})^T$ is given by
\begin{align}
M_N
&=
\left[\begin{array}{ccc}
0 & m_D^T & 0  \\ 
m_D & 0 & M^* \\ 
0  & M^* & \mu_L^* \\ 
\end{array}\right],
\end{align}
where $m_D\equiv {y_D v_3}/\sqrt2$, $M^\dag=M^*$.
In our model $\mu_L$ is given at one-loop level in fig.~\ref{fig:mu_mass}, and is explicitly computed by
\begin{align}
\mu_{L_{ij}}=& \frac{2g_{i\alpha} M_{\chi_{a}} g^T_{\alpha j}}{(4\pi)^2(M_{\chi_{a}}^2-m^2_R)(M_{\chi_{a}}^2-m^2_I)} \nonumber \\
& \times \left[M_{\chi_{a}}^2m^2_R\ln\left(\frac{M_{\chi_{a}}^2}{m^2_R}\right) 
-M_{\chi_{a}}^2m^2_I\ln\left(\frac{M_{\chi_{a}}^2}{m^2_I}\right) 
+m_I^2 m^2_R\ln\left(\frac{m_{R}^2}{m^2_I}\right) \right],
\end{align}
where $M_{\chi}\equiv y_\varphi v_\varphi/\sqrt2$, $m_{R/I}$ is the mass of $\phi^0_{4R/4I}$ which comes from real/imaginary part of $\phi^0_4$.
It implies a tiny mass scale of $\mu_L$ is expected due to the one-loop effect.
Furthermore, the mass scale of $m_D$ is of the order 1 GeV since it is proportional to $v_3$, while $M$ can be of the order 1 TeV because of bare mass.
Thus we achieve natural hierarchies among the neutral fermion mass matrices;
\begin{align}
\mu_L<< m_D<M.
\end{align}
The neutral fermions are diagonalized by a unitary matrix as follows~\cite{Kajiyama:2012xg}:
\begin{align}
V (O M_N O^T) V^T &\approx
V \left[\begin{array}{cc}
-2 (m_D^* M^{*-1}\mu^*_L M^{*-1} m_D^\dag)_{3\times3} & 0_{3\times6}   \\ 
0_{6\times3}^T & M'_{6\times6} \\ 
\end{array}\right] V^T\nn\\
&\approx  {\rm Diag}\left(D_{\nu_{1,2,3}}\ ,\ M^* - \frac{\mu^*_L}2 \ , \ M^* + \frac{\mu^*_L}2 \right), \\
&M'\equiv 
\left[\begin{array}{cc}
0 & M^*  \\ 
M^* & \mu^*_L \\ 
\end{array}\right],\\
&U=VO\approx
\left[\begin{array}{cc}
U_{\rm MNS}^{3\times3} & 0_{3\times6}   \\ 
0_{6\times3}^T & \Omega_{6\times6} \\ 
\end{array}\right]
\left[\begin{array}{cc}
1_{3\times3} & -\theta_{3\times6}   \\ 
\theta_{6\times3}^T & 1_{6\times6} \\ 
\end{array}\right],\\
&
\theta_{3\times6}\approx [-m_D^* M^{*-1}\mu^\dag_L M^{*-1} , m_D^* M^{*-1}],\\
&
\Omega_{6\times6}\approx\frac{1}{\sqrt2} 
\left[\begin{array}{cc}
i\left(1+\frac{\mu_L M^{*-1}}{4}\right) & -i\left(1-\frac{\mu_L M^{*-1}}{4}\right)   \\ 
1-\frac{M^{*-1} \mu_L^\dag }{4} &1+\frac{\mu_L^* M^{*-1}}{4} \\ 
\end{array}\right],
\end{align}
where $\Psi_L=U^T N_L$, $N_L$ being mass eigenstates with nine components, and $M\pm\mu_L/2\approx {\rm Diag}(M\pm\mu_L/2)$. 

\noindent \underline{\it  Active neutrino sector}:
Here let us focus on the active neutrino sector; $m_\nu\equiv -2 m_D^* M^{*-1}\mu^*_L M^{*-1} m_D^\dag=U_{\rm MNS}^\dag D_\nu U_{\rm MNS}^*$  ($U_{\rm MNS} \equiv U_{\rm MNS}^{3\times3}$) from the above definition.
Here since we define $\mu\equiv  M^{*-1}\mu^*_L M^{*-1}$, where $\mu$ is symmetric matrix.
Then one rewrites $\mu\equiv R R^T$, where $R$ is triangular matrix and $R$ is uniquely given by each the component of $\mu$; $m_\nu=-(\sqrt{2}m^*_DR)(\sqrt{2}R^T m_D^T)\equiv -rr^T$. Applying Casas-Ibarra parametrization~\cite{Casas:2001sr}, one finds the following relation:
\begin{align}
&y_D=i\frac{U^\dag_{\rm MNS} \sqrt{D_\nu} {\cal O} R^{-1}}{v_3}\lesssim \sqrt{4\pi},
\quad 
R^{-1}=
\left[\begin{array}{ccc}
\frac1a & 0 & 0  \\ 
-\frac{d}{ab} & \frac{1}{b} & 0 \\ 
\frac{-be+df}{abc}  & \frac{f}{bc} &\frac{1}{c}  \\ 
\end{array}\right],\\
&
a=\sqrt{\mu_{11}},\quad d=\frac{\mu_{12}}{a},\quad b=\sqrt{\mu_{22}-d^2},\\
&
e=\frac{\mu_{13}}{a},\quad f=\frac{\mu_{32}-de}{b},\quad c=\sqrt{\mu_{33}-e^2-f^2},
\end{align}
where ${\cal O}$ is an arbitrary three by three orthogonal matrix with complex values; ${\cal O}{\cal O}^T={\cal O}^T{\cal O}=1$.

 \subsection{Constraints from running of gauge coupling and LFV}

\begin{figure}[tb]
\begin{center}
\includegraphics[width=10.0cm]{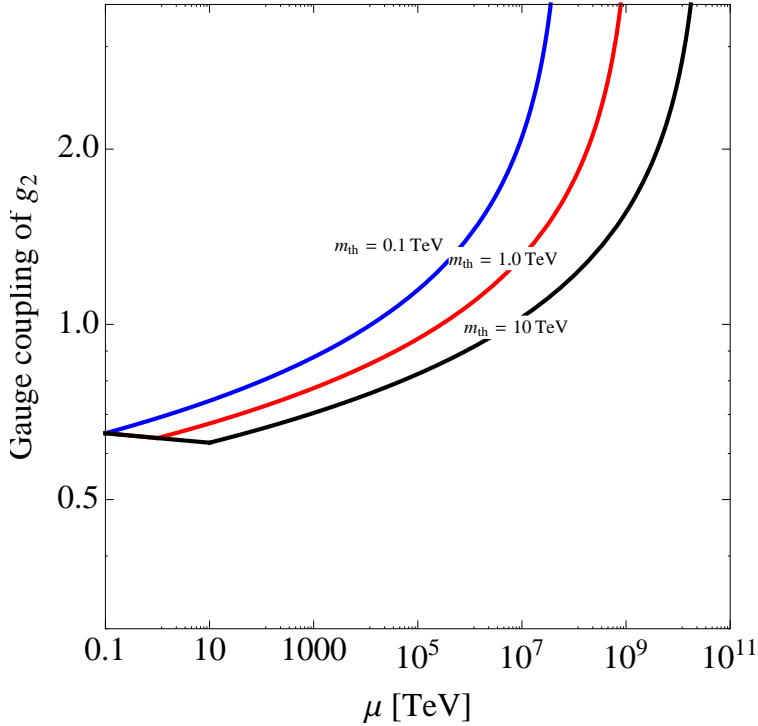}
\caption{The running of $g_2$ in terms of a reference energy of $\mu$, where the black line corresponds to  $m_{th}=$10 TeV,
the red line corresponds to  $m_{th}=$1.0 TeV, and the blue one does  $m_{th}=$0.1 TeV. }
\label{fig:rge}
\end{center}\end{figure}

\noindent \underline{\it Beta function of $SU(2)_L$ gauge coupling $g_2$}:
\label{beta-func}
Here we discuss the running of gauge coupling of $g_2$.
The new contribution to $g_2$ for a $SU(2)_L$ quartet fermion(boson) $\psi(H_4)$, and a triplet boson $H_3$,are respectively given by
\begin{align}
 \Delta b^{\psi}_{g_2}=\frac{10}{3}, \ \Delta b^{H_3}_{g_2}=\frac{2}{3},\ \Delta b^{H_4}_{g_2}=\frac{5}{3}  .
\end{align}
Then one finds the energy evolution of the gauge coupling $g_2$ as~\cite{Nomura:2017abu, Kanemura:2015bli}
\begin{align}
&\frac{1}{g^2_{g_2}(\mu)}=\frac1{g^2(m_{in})}-\frac{b^{SM}_{g_2}}{(4\pi)^2}\ln\left[\frac{\mu^2}{m_{in}^2}\right]
-\theta(\mu-m_{th}) 
 \frac{N_{f_\psi} \Delta b^{\psi}_{g_2} +\Delta b^{H_3}_{g_2}+\Delta b^{H_4}_{g_2}}{(4\pi)^2}\ln\left[\frac{\mu^2}{m_{th}^2}\right],
\label{eq:rge_g}
\end{align}
where $N_{f_{\psi}}=3$ is the number of $\psi$, $\mu$ is a reference energy, $b^{SM}_{g_2}=-19/6$, and we assume to be $m_{in}(=m_Z) < m_{th}$, being $m_{th}$ threshold masses of exotic fermions and bosons.
The resulting flow of ${g_2}(\mu)$ is then given by the Fig.~\ref{fig:rge}.
This figure shows that the black line is relevant up to the mass scale $\mu={\cal O}(10^{11})$ TeV in case of $m_{th}=$10 TeV,
the red one is relevant up to the mass scale $\mu={\cal O}(10^{9})$ TeV in case of $m_{th}=$1 TeV,
and the blue one is relevant up to the mass scale $\mu={\cal O}(10^7)$ PeV in case of $m_{th}=$0.1 TeV.
Thus our theory is valid up to the typical energy scale of grand unified theory (GUT) $\sim10^{15}$ GeV.

\noindent \underline{\it  Lepton flavor violations(LFVs)}:
LFVs arise from the term $f$ at one-loop level, and its form can be given by~\cite{Lindner:2016bgg, Baek:2016kud}
\begin{align}
 {\rm BR}(\ell_i\to\ell_j\gamma)= \frac{48\pi^3\alpha_{\rm em} C_{ij} }{{\rm G_F^2} m_{\ell_i}^2}\left(|a_{R_{ij}}|^2+|a_{L_{ij}}|^2\right),
 \end{align}
where 
 \begin{align}
 a_{R_{ij}} &=\frac{m_{\ell_i}} {(4\pi)^2}
\left[\sum_{\alpha=1}^9\frac{Y_{D_{j\alpha}} Y^\dag_{D_{\alpha i}}}3 F(\psi^0_\alpha,\delta^-)
-\sum_{\beta=1}^3
y_{D_{j\beta}} y^{\dag}_{D_{\beta i}} [ 2 F(\delta^-, \psi^{--}_\beta) + F(\psi^{--}_\beta,\delta^-) +\frac13  F(\delta^0, \psi^{-}_\beta)]  
\right],
\end{align}
$a_L=a_R(m_{\ell_i}\to m_{\ell_j})$, $Y_{D_{i\alpha}} \equiv \sum_{j=1}^3 y_{D_{ij}} U^T_{j+3,\alpha}$, and 
\begin{align}
&F(a,b)\equiv \frac{1}{2m_a^2}\int_0^1dx \frac{x(1-x)^2}{x+(1-x)r_{ab}},\quad r_{ab}\equiv\frac{m_b^2}{m_a^2}.
\end{align}

\noindent \underline{\it New contributions to the muon anomalous magnetic moment (muon $g-2$: $\Delta a_\mu$)}: 
In the model $\Delta a_\mu$ arises from the same terms of LFVs and can be formulated by the following expression:
Also another source via additional gauge sector can also be induced by
\begin{align}
&\Delta a_\mu \approx -m_\mu [{a_{L_{\mu\mu}}+a_{R_{\mu\mu}}}] 
= -2m_\mu{a_{L_{\mu\mu}}}, \label{eq:G2-ZP}
\end{align}
where we use the allowed range $\Delta a_\mu= (26.1\pm16.0)\times 10^{-10}$~\cite{Hagiwara:2011af} (at 2$\sigma$ C.L.) in our numerical analysis below.

\subsection{Dark Matter}
\begin{figure}[tb]
\begin{center}
\includegraphics[width=10.0cm]{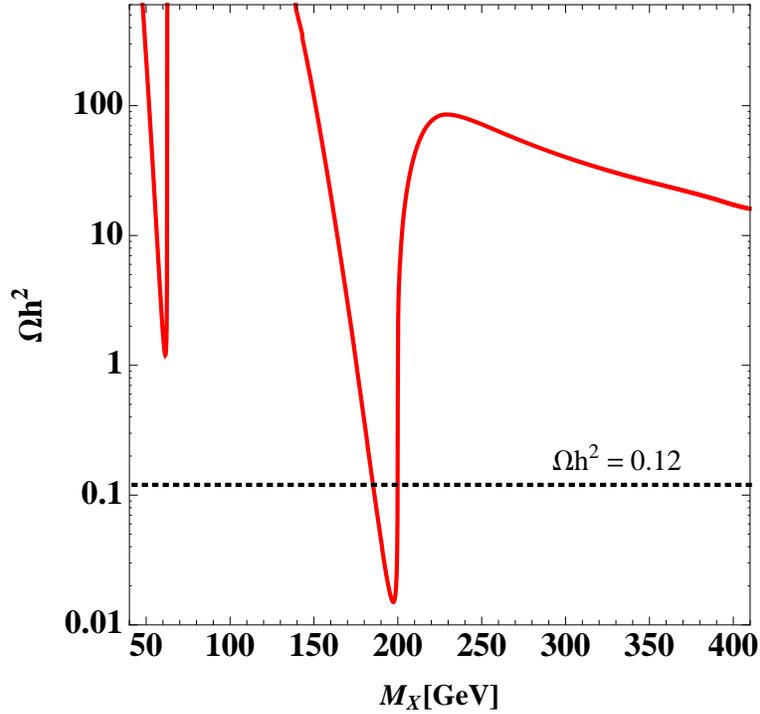}
\caption{Relic density of DM as a function of DM mss fixing $m_\phi = 400$ GeV, $v_\varphi = 2000$ GeV and $\sin \alpha = 0.1$}
\label{fig:DM}
\end{center}\end{figure}

 Here we briefly discuss the feature of our DM candidate, which is assumed to be the lightest $Z_2$ odd Majorana fermion $X \equiv \chi_R$; $M{\chi_{R_1}}\equiv M_X$.
The relevant interactions are given by 
\begin{equation}
\mathcal{L} \supset \frac{M_X}{2 v_\varphi} \cos \alpha \phi \bar X^c X + \frac{M_X}{2 v_\varphi} \sin \alpha h \bar X^c X + i \frac{M_X}{2 v_\varphi} a_\varphi \bar X^c \gamma_5 X,
\end{equation}
where $M_X$ is DM mass given by $M_X \equiv M_1 = y_{\varphi_{11}} v_\varphi/\sqrt{2}$. 
Also we have interactions among $\phi$ and SM particles whose couplings are obtained by putting $-\sin \alpha$ to SM Higgs couplings. 
Then dominant DM annihilation processes are 
\begin{align}
& XX \to \phi/h \to \bar f_{SM} \bar f_{SM}/ZZ/W^+W^-, \\
& XX \to a_\varphi a_\varphi/a_\varphi \phi(h)
\end{align}
where the first modes are s-channel via $\varphi$-$H$ mixing and the second ones are t and u channels with physical GB final state. 
In Fig.~\ref{fig:DM}, we show the relic density of $X$ as a function of its mass fixing $m_\phi = 400$ GeV, $v_\varphi = 2000$ GeV and $\sin \alpha = 0.1$, which is estimated by 
{\it micrOMEGAs 4.3}~\cite{Belanger:2014vza} implementing relevant interactions.
We find that relic density can be explained around $2 M_X \sim m_\phi$ by resonant enhancement of annihilation cross section; we also see resonance effect at $2 M_X \sim m_h$ but relic density is too large in this parameter setting.  
The relic density also decreases as $M_X$ increases since scalar interactions are proportional to $M_X$.
In addition, relic density can be explained via GB mode when $v_\varphi$ is smaller
as can be seen in e.g ref.~\cite{Nomura:2017wxf}.
The  resonant point is consistent with the constraint of direct detection since DM-scalar coupling can be small~\cite{Kanemura:2010sh}.
But it depends on the parameter space for solution without resonance.

In case of the bosonic DM candidate that is not considered as a DM one in our whole analysis,
the dominant contribution to the relic density comes from kinetic terms, and the free parameter is almost  the DM mass only.
Thus the formulation is already established by ref.~\cite{Cirelli:2005uq, Cirelli:2007xd}, which suggests $M_X\sim$ 10 TeV.
As a result, any masses of new fields must be 10 TeV or larger than 10 TeV, and one cannot detect any new particles at current colliders.

\section{Numerical analyses}

In this section, we carry out numerical analysis taking into account neutrino mass and LFV constraints 
exploring possible value of Yukawa coupling $y_D$ and $\Delta a_\mu$. 
In addition collider physics is discussed focusing on doubly charged lepton production at the LHC.

\subsection{Yukawa coupling and $\Delta a_\mu$ in benchmark points}
Here we have  numerical analysis in two benchmark points, where we commonly fix the following values:
\begin{align}
&\theta_{23}\approx 0.62+1.08i,\quad 
\theta_{13}\approx 0.46+0.69i,\quad 
\theta_{12}\approx 1.82+14.95i,\\
&  m_{\nu_1}=0.1\ {\rm meV}, \quad v_3\approx 3\ {\rm GeV},\\
& g\approx
\left[\begin{array}{ccc}
0.2 & 0.027 & 0.000020  \\ 
0.000021 & 0.091 & 0.083 \\ 
0.0033  & 0.000034 & 0.58 \\ 
\end{array}\right],
 \end{align}
 where $\theta_{12,13,23}$ is the mixings of ${\cal O}$ introduced in the analysis of neutrino mass matrix,
 and these values; especially $\theta_{12,13,23}$, are selected so as to maximize muon $g-2$ while minimizing  LFVs. \\
\noindent \underline{\it  Bench mark point 1}:
The first bench mark point is given by  
\begin{align}
& (M_{\chi_2},M_{\chi_3})=(2,2.5)\ {\rm TeV},\quad (M_{1},M_2,M_3)=(0.4,0.45,0.5)\ {\rm TeV},
\nn\\ 
&(m_R,m_I)=(3,3.3)\ {\rm TeV},\quad (m_{\delta^0},m_{\delta^\pm})=(0.3, 0.35)\ {\rm TeV}.
\end{align}
\noindent \underline{\it  Bench mark point 2}:
The second bench mark point is given by  
\begin{align}
& (M_{\chi_2},M_{\chi_3})=(4,4.5)\ {\rm TeV},\quad (M_{1},M_2,M_3)=(0.5,0.55,0.6)\ {\rm TeV},
\nn\\ 
&(m_R,m_I)=(5,5.5)\ {\rm TeV},\quad (m_{\delta^0},m_{\delta^\pm})=(0.4, 0.45)\ {\rm TeV}.
\end{align}

\begin{figure}[tb]\begin{center}
\includegraphics[width=8cm]{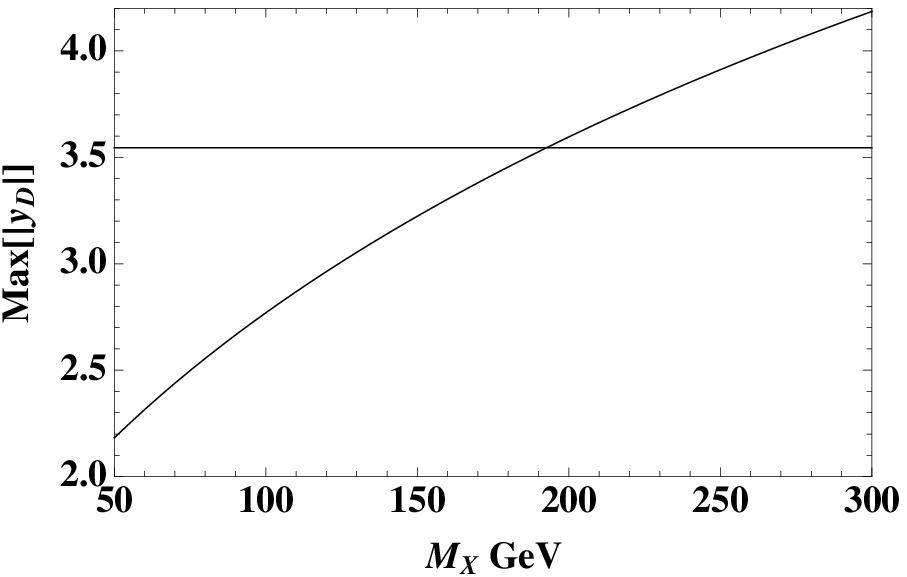}
\includegraphics[width=8cm]{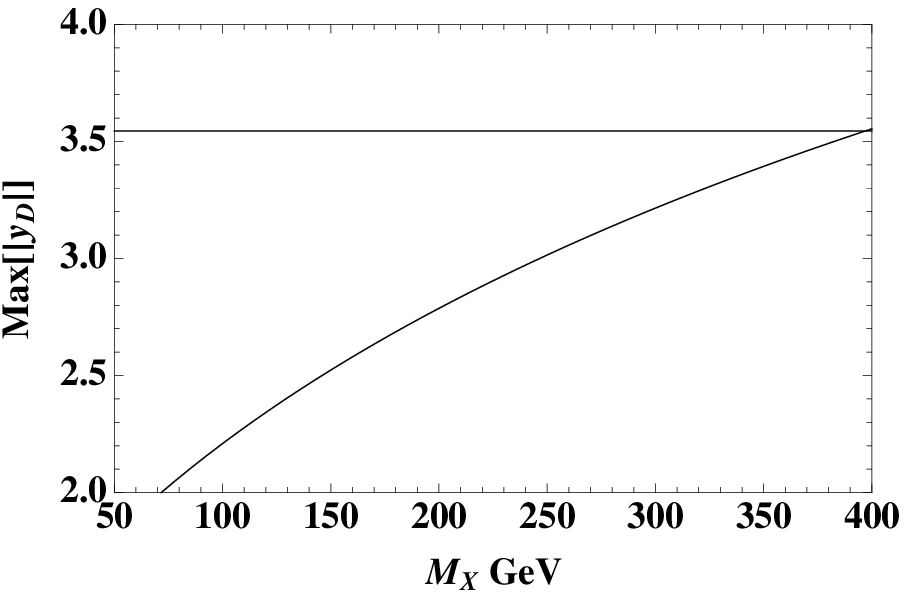}
\caption{A line of the maximum absolute value of $y_D$(=Max[$|y_D|$]) in terms of  the DM mass $M_X\equiv M_{\chi_1}$,
where the left(right)-hand side shows the bench mark point 1(2), and the horizontal line represents the perturbation limit; $\sqrt{4\pi}$.
}   \label{fig:yd}
\end{center}\end{figure}

\begin{figure}[tb]\begin{center}
\includegraphics[width=8cm]{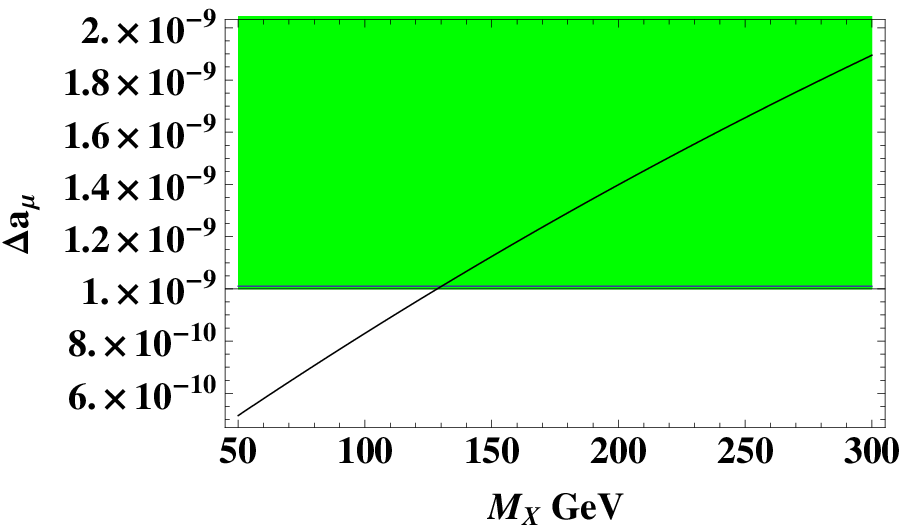}
\includegraphics[width=8cm]{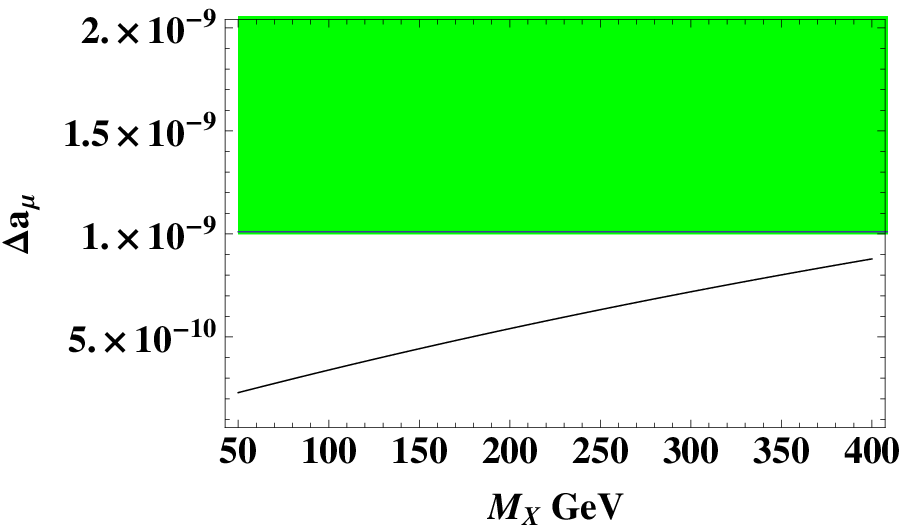}
\caption{Line of $\Delta a_\mu$ in terms of  $M_{X}$, where the left(right)-hand side shows the bench mark point 1(2).
The green region is expected  to be detected by the experiment that covers over $\Delta a_\mu= (26.1\pm16.0)\times 10^{-10}$ at 2$\sigma$ C.L..
}   \label{fig:lfvs}\end{center}\end{figure}

Fig.~\ref{fig:yd} represents the flow of maximum component of $y_D$(=Max[$|y_D|$]) in terms of  the DM mass $M_X\equiv M_{\chi_1}$,
where the left(right)-hand side shows the bench mark point 1(2), and the horizontal line represents the perturbation limit; $\sqrt{4\pi}$.
The left(right)-hand side figure suggests that the allowed region of DM mass be less than $\sim$ 190(400) GeV, satisfying the perturbative limit.
Fig.~\ref{fig:lfvs} represents the flow of maximum component of $\Delta a_\mu$ in terms of $M_{X}$, where the left(right)-hand side shows the bench mark point 1(2).
The green region is expected  to be detected by the experiment that covers over $\Delta a_\mu= (26.1\pm16.0)\times 10^{-10}$ at 2$\sigma$ C.L..
It implies that the left-bench mark point can reach the observed region of muon $g-2$ at 130 GeV$\lesssim M_X\lesssim$190 GeV,
while the right one is below the allowed region of  muon $g-2$ in the perturbative limit even though the allowed region of DM mass is wider than the left one.

\subsection{collider physics}
Here we briefly discuss collider signature of the model. 
There are several exotic charged particles in the model coming from $SU(2)_L$ multiplet fermions and scalar fields.
Interestingly we have two doubly charged particles in fermion and scalar sector.
The doubly charged lepton $\psi^{--}$ in quartet fermion can provide interesting signal at the LHC since it decays into SM particles and its mass could be reconstructed.
On the other hand doubly charged scalar $\phi_4^{++}$ in $H_4$ decays into final state including DM due to $Z_2$ oddness, and mass reconstruction is more difficult.
We thus concentrate on doubly charged lepton production and its decay at the LHC.

Firstly doubly charged lepton can be produced via electroweak interactions where we consider pair production process.
Relevant gauge interactions are given by
\begin{equation}
\mathcal{L} \supset - 2 e A_\mu \bar \psi^{--} \gamma^\mu \psi^{--} + \frac{g_2}{c_W} \left( -\frac32 +2 s_W^2 \right) Z_\mu \bar \psi^{--} \gamma^\mu \psi^{--}
\end{equation} 
where $s_W(c_W) = \sin \theta_W (\cos \theta_W)$ with Weinberg angle $\theta_W$ and $e$ is electromagnetic coupling.
Taking the first generation mass as $M_1 = 500$ GeV as a benchmark point we obtain pair production cross section $\sigma(pp \to \psi^{++}_1 \psi^{--}_1)$ 
around $40$ fb estimated by {\it CalcHEP}~\cite{Belyaev:2012qa}; here we focus on first generation $\psi_{1}^{\pm \pm}$ since it provides the largest production cross section and omit generation index below.
Then $\psi^{\pm \pm}$ decays into charged lepton and singly charged scalar $\delta^\pm$ through the Yukawa interaction in Eq.~(\ref{Eq:yuk})
where we write by components such that
\begin{align}
& y_{D_{ab}} [\bar L^a_L H_3 \psi_R^b] \nonumber \\ 
& = \frac{y_{D_{ab}}}{\sqrt{3}} \left[ \bar e^a_L \left( \psi_R^0 \delta'^- - \sqrt{2} \psi'^-_R \delta^0 + \sqrt{3} \psi^{--}_R \delta^+ \right) 
+ \bar \nu_L \left( \psi'^-_R \delta^+ - \sqrt{2} \psi^0_R \delta^0 + \sqrt{3} \psi^+_R \delta'^- \right) \right].
\end{align}
For simplicity we assume $\ell= e, \mu$ in the decay of $\psi^{--} \to \ell^- \delta^-$ in the following discussion.
The singly charged scalar $\delta^\pm$ dominantly decay into $W^+ Z$ since $H_3$ do not couple to SM fermions directly.
In total we have following signal process: 
\begin{equation}
pp \to \psi^{++} \psi^{--} \to \delta^+ \delta^- \ell^+_1 \ell^+_2 \to W^+W^-ZZ \ell^+_1 \ell^-_2.
\end{equation}
Taking into account decays of SM gauge bosons our signal at the LHC will be multi-lepton with or without jets.
In Table~\ref{tab:2}, we summarize products of cross section and branching ratios (BRs) for each representative final state 
where we take doubly charged lepton mass as $M = 500$ GeV as a benchmark point and final states are distinguished by number of charged lepton and jets.
We find that the cross section is less than 1 fb when there are more than 4 charged leptons in final state.
Thus sufficiently large integrated luminosity is required to analyze the signal at the LHC experiments.
Detailed simulation study including SM backgrounds is beyond the scope of this letter and it will be given elsewhere.

\begin{table}[t!]
\begin{tabular}{|c||cccccc|}\hline\hline  
Signal & $4 \ell^+ 4 \ell^- \slashed{E}_T$~ & ~$3 \ell^+ 3 \ell^- 2j \slashed{E}_T$~ & ~$2 \ell^+ 2\ell^- 4j \slashed{E}_T$~ & ~$4 \ell^\pm 3 \ell^\mp 2j \slashed{E}_T$~ &  ~$3 \ell^\pm 2 \ell^\mp 4j \slashed{E}_T$~ 
& ~$3 \ell^+ 3 \ell^- 4j $~\\\hline\hline 
$\sigma \cdot BR$ [fb] & 0.0097 & 0.19 & 0.97 & 0.053 & 0.053 & 0.072 \\ \hline 
\end{tabular}
\caption{ Some final states from doubly charged lepton pair production with values of $\sigma(pp \to \psi^{++} \psi^{--}) BR$ fixing $M_1 = 500$ GeV.}
\label{tab:2}
\end{table}

\section{Summary and discussion}
In this work, we constructed inverse seesaw model with global lepton number symmetry
in which Majorana mass term of extra neutral fermion from $SU(2)_L$ quartet is induced at one-loop level realizing natural hierarchy of neutral fermion masses.
In our model, $Z_2$ odd scalar quartet and fermion singlet are introduced which propagate inside a loop diagram generating the Majorana mass.
Then the lightest $Z_2$ odd particle can be a good DM candidate where we consider the Majorana fermion DM.

We have formulated neutrino mass matrix, LFV and muon $g-2$. 
In addition we show that relic density can be explained by scalar exchanging interactions.
Then numerical analysis is carried out exploring allowed value of coupling constant and muon $g-2$.
We find that sizable muon $g-2$ can be obtained when DM mass is 130 GeV $\leq M_X \leq$ 190 GeV taking into account perturbative limit of coupling constants.   
Furthermore we have discussed collider physics focusing on doubly charged lepton production at the LHC
where we have shown the products of cross section and branching ratio for each final state.
We could test the signals with sufficiently large integrated luminosity.

{Finally we discuss possibility of embedding this model in a grand unified theory such as an $SO(10)$ model~\cite{Malinsky:2005bi}. 
In our model, we have introduced $SU(2)_L$ quartet fermions and scalar field to realize one loop generation of Majorana mass term in inverse seesaw,
and a large $SO(10)$ multiplet is required to obtain $SU(2)_L$ quartet such as ${\bf 210'}$ or ${\bf 320}$ representation~\cite{Slansky:1981yr}.
Thus it is non-trivial to embed our model in a $SO(10)$ model and further investigation is left for future study.
}

\section*{Acknowledgments}
This research is supported by the Ministry of Science, ICT and Future Planning, Gyeongsangbuk-do and Pohang City (H.O.). 
And H. O. is sincerely grateful for KIAS and all the members, too.

\end{document}